# A Comparative Study of Various Routing Protocols in VANET

[1]Rakesh Kumar, [2]Mayank Dave
[1] Department of IT, M. M. University, Mullana, Haryana, India

[2] Department of Computer Engineering, N.I.T. Kurukshetra, Haryana, India

**Abstract**

Vehicular Ad Hoc Networks (VANET) is a subclass of Mobile ad hoc networks which provides a distinguished approach for Intelligent Transport System (ITS). The survey of routing protocols in VANET is important and necessary for smart ITS. This paper discusses the advantages / disadvantages and the applications of various routing protocols for vehicular ad hoc networks. It explores the motivation behind the designed, and traces the evolution of these routing protocols. Finally the paper concludes by a tabular comparison of the various routing protocols for VANET.

## 1. Introduction

Vehicular networks represent a particularly new class of wireless ad hoc networks that enable vehicles to communicate with each other and/or with roadside infrastructure. Earlier, drivers were using their voice, gestures, horns, and observation of each other's trajectory to manage their behavior. When the drastic increase of vehicles made this not enough to manage, in the second half of the 19th century, traffic police took charge of controlling and managing the traffic using hand signals, semaphores and colored lights. The 1930s saw the automation of traffic signals and in the 1940s car indicators were deployed widely. Variable-message signs were introduced in the 1960s to provide information to the drivers to adapt according to the current circumstances. The information communicated via all of these means is, however, very less: road infrastructure typically provides the similar information to all cars, and the amount of information that the drivers can share directly with one another is restricted. Recently, drivers can exchange more information, such as traffic information and directions, to each other via car phones or citizen band radio.

Wireless communication supports more customized and complete information to be exchanged. VANET addresses all these issues related to the communications between vehicles and on-going research with wireless communication. It also covers the aspects of Wireless Access for the Vehicular
Environment (WAVE) standards based on the emerging IEEE 802.11p specification. VANET basically enables *infrastructure-to-vehicle (I2V)*, *vehicle-to-infrastructure (V2I)*, and *vehicle-to-vehicle (V2V)* communications. In this paper, we use the term V2I to refer to both I2V and V2I communication.

## 2. Network Architectures

Wireless ad hoc networks generally do not rely on fixed infrastructure for communication and dissemination of information. VANETs follow the same principle and apply it to the highly dynamic environment of surface transportation. As shown in Fig. 1, the architecture of VANETs mainly falls within three categories: pure cellular/WLAN, pure ad hoc, and hybrid. VANETs may use fixed cellular gateways and WLAN / WiMax access points at traffic intersections to connect to the Internet, gather traffic information, or for routing purposes. The network architecture under this scenario is a pure cellular or WLAN structure as shown in Fig. 1(a). VANETs can combine both cellular network and WLAN to form the networks so that a WLAN is used where an access point is available and a 3G connection otherwise.

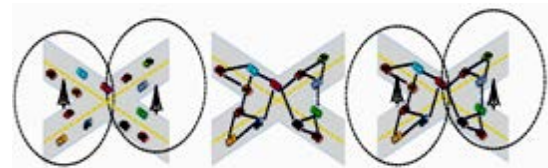

**(a) Cellular/WLAN  (b) Ad Hoc  (c) Hybrid**

**Figure 1:** Network architectures for VANETs

Stationary or fixed gateways around the sides of roads could provide connectivity to mobile nodes (vehicles), but are eventually unfeasible considering the infrastructure costs involved. In such a scenario, all vehicles and road-side wireless devices can form a pure mobile ad hoc network (Fig. 1(b)) to perform vehicle to vehicle communications and achieve certain goals, such as blind crossing.

Hybrid architecture (Fig. 1(c)) of combining infrastructure networks and ad hoc networks together has also been a possible solution for VANETs. Namboodiri et al. [13] proposed such a hybrid architecture, which uses some vehicles with both WLAN and cellular capabilities as the gateways and mobile network routers so that vehicles with only WLAN capability can communicate with them through multi-hop links to remain connected to the world. The hybrid architecture can provide better coverage, but also causes new problems, such as the seamless transition of the communication among different wireless systems.

## 3. Layered View Of Vehicular Networks

Vehicular networks can be classified on the basis of five different aspects as shown in table 1. Vehicular Networks has the diverse range of applications that varies safety applications to comfort applications.

**Table 1**: Layered View of vehicular networks

| Vehicular Network | Application Type | • Safety application<br>• Intelligent transport application<br>• Comfort application |
|---|---|---|
| | Quality of Service | • Non-real-time<br>• Soft-real-time<br>• Hard-real-time |
| | Scope | • Wide area<br>• Local |
| | Network Type | • Ad hoc<br>• Infrastructure-based |
| | Communication Type | • V2I<br>• V2V |

Safety Applications enhances the driving conditions and reduces the chances of accidents by providing enough time to the driver and applying the brakes automatically. These can be further divide into the following:

- Cooperative collision warning
- Incident management
- Emergency video streaming

Intelligent transport applications aim at providing faster delivery of traffic information, and improving the efficiency and accuracy of traffic detection by allowing collaborative processing of information between vehicles. These applications focus on observing the traffic pattern and managing traffic accordingly. It can be further categorized into the following:

- Traffic Monitoring
- Traffic Management
- Platooning
- Vehicle tracking
- Notification services

Comfort applications are the applications of VANET related to comfort level of the passenger moving in the vehicle. It can be further categorized into the following:

- Parking place management.
- Distributed games and/or talks.
- Peer to Peer applications

Consequently, the Quality of Service (QoS) required for the network varies from *non-real-time*, to *soft real-time* where a timing failure might compromise service quality, up to *hard real-time* where a timing failure might lead to a catastrophe. These applications can also be exemplified by their scope, i.e., whether they provide communication over a *wide area*, or are *local* only. Finally, such applications can vary in their networking approach: *ad hoc*, where vehicles communicate suddenly, or *infrastructure-based*, where communication is governed by fixed base stations. VANET has the communication type: Vehicle to Vehicle (*V2V*) and Vehicle to Infrastructure (*V2I*).

## 4. VANET Characteristics, Issues And Requirements

Wireless communication, particularly real-time communication is highly unreliable. In addition, VANET has certain unique issues that make it different from other wireless networks. Because no central coordination can be assumed, a sole shared control channel is required at the MAC layer (the so-called one channel paradigm). Mobility movements of vehicular networks are also very specific, e.g. vehicles move along the roads, in predefined directions, and this requires new specific mobility models to be devised. Normal mobility models could not address the requirements of VANET. Moreover, now a day's cars are having very high mobility rates and so change the topology in an in-deterministic fashion that makes wireless transmission very challenging.

Furthermore, the vehicle density exhibits spatio-temporal variations: it m ight be very scarce (eg. Highway), with no vehicle or only few, up to very dense (eg. city area), with over 500 vehicles per kilometer. Both ends of the density spectrum are particularly challenging. The applications of vehicular networks should also fulfill a number of nonfunctional requirements, such as potentially very high reliability, but also security to ensure that safety-critical applications cannot be tempered with. Vehicles range over very large geographical areas (cities or countries), and therefore require potentially large-scale networks, and especially a v ery extensive deployment of equipment if infrastructure-based networks are used. Many VANET applications have either delay constraints or other QoS requirements. Efficient broadcasting of safety messages for getting full coverage and low latency to provide QoS and reliability in VANET routing is still a challenging problem[8].

Since mobility of VANETs cannot be captured by general mobility models. Traffic flow (both in time and space) need to be studied and integrated in the design of reliable and high-performance mobility models.

Apart from this security is also one of the major issues in VANET. Cooperation among inter-vehicular networks and sensor networks placed within the vehicles or along the road need to be further investigated and analyzed. As the number of vehicles grows the trust between them should also be maintained for the smooth communication.

In addition to technical challenges, socio-economic challenges have to be solved. The benefits of V2V communication only become significant when there are a s ufficiently large number of vehicles using the technology. Vehicular applications must therefore be able to operate and be useful despite initial low penetration.

## 5. Overview Of Routing Protocols

In VANET, the routing protocols are classified into five categories: Topology based routing protocol, Position based routing protocol, Cluster based routing protocol, Geo cast routing protocol and Broadcast routing protocol. These protocols are characterized on the basis of area / application where they are most suitable [1].

### a) Topology Based Routing Protocols

These routing protocols use links information that exists in the network to perform packet forwarding. They are further divided into Proactive and Reactive.

#### i) Proactive routing protocols

The proactive routing means that the routing information, like next forwarding hop is maintained in the background irrespective of communication requests. The advantage of proactive routing protocol is that there is no route discovery since the destination route is stored in the background, but the disadvantage of this protocol is that it provides low latency for real time application. A table is constructed and maintained within a n ode. So that, each entry in the table indicates the next hop node towards a certain destination. It also leads to the maintenance of unused data paths, which causes the reduction in the available bandwidth. T he various types of proactive routing protocols are: LSR, FSR.

#### ii) Reactive/Ad hoc based routing

Reactive routing opens the route only when it is necessary for a n ode to communicate with each other. It maintains only the routes that are currently in use, as a r esult it reduces the burden in the network. R eactive routing consists of route discovery phase in which the query packets are flooded into the network for the path search and this phase completes when route is found. T he various types of reactive routing protocols are AODV, PGB, DSR and TORA

### b) Position Based Routing Protocols

Position based routing consists of class of routing algorithm. They share the property of using geographic positioning information in order to select the next forwarding hops. The packet is send without any map knowledge to the one hop neighbor, which is closest to destination. Position based routing is beneficial since no global route from source node to destination node need to be created and maintained. Position based routing is broadly divided in two types: Position based greedy V2V protocols, Delay Tolerant Protocols.

#### 1) Position Based Greedy V2V Protocols

In greedy strategy and intermediate node in the route forward message to the farthest neighbor in the direction of the next destination. Greedy approach requires that intermediate node should possessed position of itself, position of its neighbor and destination position. The goal of these protocols is to transmit data packets to destination as soon as possible that is why these are also known as min delay routing protocols. V arious types of position based greedy V2V protocols are GPCR, CAR and DIR

2) Greedy Perimeter Coordinator Routing (GPCR)

GPCR is based upon the fact that city street form a natural planner graph. GPCR does not require external static street map for its operation. GPCR consists of two components: A Restricted Greedy forwarding procedure, A repair strategy for routing algorithm. A GPCR follows a destination based greedy forwarding strategy, it routes messages to nodes at intersection. Since GPCR does not use any external static street map so nodes at intersection are difficult to find. GPCR uses heuristic method for finding nodes located at intersections and designates those nodes as coordinators. Coordinator has the responsibility of making routing decisions. There are two approaches used for coordinator determination they are (a) Neighbor Table Approach: The nodes periodically transmit beacon messages which contains their position information and last known position information of all neighbors, by listening to beacon messages a node as information about its own position, position of its neighbor and neighbor's neighbor. Using this information node X consider itself to be within the intersection. (b) Correlation coefficient approach: In this case node uses its position information and the position information of its immediate neighbor to find the correlation coefficient, $p_{xy}$. This approach performs better than neighbor table approach. By using this approach the algorithm can avoid dependencies on external street map.

3) Connectivity Aware Routing Protocols (CAR)

CAR protocols find a route to a destination; it has unique characteristics that it maintains the cache of successful route between various source and destination pairs. It also predicts the position of destination vehicle repairs route as the position changes. Nodes using CAR protocols send periodic Hello beacons that contain their velocity vector information. On receiving Hello beacons a node will record sender in its neighbor table and calculate its own velocity vector and velocity vector of its neighbor. Beacons can also be piggybacked on forwarded data packets to reduce wastage of bandwidth and congestion. Entries expire from the neighbor table when the distance between nodes exceeds the threshold value. The CAR protocols establishes the notation of a guard which is a geographic marker message, it is buffered and passed from one vehicle to another to propagate the information. A guard is a temporary message that has an ID, a TTL (Time to live) counts, a radius and some state information. CAR provides two forms of guards. The Standing guard and The Traveling guard. Routing errors may occur due to communication gap between anchor points or due to guards. So CAR protocol has two recovery strategies to cope with the problem. The first strategy is Time out algorithm with active waiting cycle. The second strategy is walk around error recovery. The CAR protocol has the ability to generate virtual information in the form of guards, which is a distinct advantage over other protocols.

4) Diagonal-Intersection-Based Routing Protocol (DIR)

DIR protocol constructs a series of diagonal intersections between the source and destination vehicle. The DIR protocol is based upon the geographic routing protocol in which source vehicle geographically forwards the data packets towards the first diagonal intersection, second diagonal intersection and so on until the last diagonal intersection and finally geographically reaches to designation vehicle. DIR vehicle is auto adjustable, Auto adjustability means that one sub path with low data packet delay between two neighboring diagonal intersections, which is dynamically selected to forward data packets. To reduce the data packet delay the route is automatically selected with lowest sub path delay. DIR protocol can automatically adjust routing path for keeping the lower packet delay.

5) Delay Tolerant Protocols

In urban scenario where vehicle are densely packed locating a node to carry a message is not a problem but in rural highway situation or in cities at night fewer vehicles are running and establishing end to end route is difficult. So in such cases certain consideration needs to be given in sparse networks. The various types of Delay Tolerant Protocols are MOVE, VADD, and SADV.

6) Motion Vector Routing Algorithm (MOVE)

The MOVE algorithm is an algorithm for sparse VANET scenario. In these scenarios vehicle act as mobile router that have intermittent connectivity with other vehicles. Connection opportunities must be scrutinized carefully since they occur infrequently and global topology is also rapidly changes. The algorithm must predict whether forwarding message will provide progress toward intended destination. MOVE algorithm assumes that each node has knowledge of its own position, heading and destination. From this information the current vehicle node can calculate the closest distance between the vehicle and message destination. M OVE algorithm use less buffer space. MOVE algorithm is specially designed for sparse networks and for vehicles that transfer data from sensor networks to base station.

7) Vehicle Assisted Data Delivery (VADD)

VADD uses a car ry and forward strategy to allow packets to be carried by vehicle in sparse networks for forwarding when the node enters the broadcast range, thereby allowing a packet to be forwarded by relay in case of sparse networks. VADD require each vehicle to know its own position and also require an external static street map. E ach packet has three modes: Intersection, StraightWay and Destination, where each mode is based on the location of the node carrying the packet. Intersection mode is used when the packet has reached an intersection at which routing decisions can be made for the packet to be forwarded to a vehicle along any of the available directions of the intersection. In StraightWay mode the current node is on a road where there are only two possible directions for the packet to travel, in the direction of the current node or in the opposite direction. Destination mode is when the packet is close to its final destination.

8) Static Node Assisted Adaptive Routing Protocol (SADV)

SADV aims at reducing message delivery delay in sparse networks. S ADV also dynamically adapts to varying traffic density by allowing each node to measure the amount of time for message delivery. S ADV assumes that each vehicle knows its position through GPS and each vehicle has accessed to external static street map. S ADV has three different modules; Static Node Assisted Routing (SNAR), Link Delay Update (LDU) and Multipath Data Dissemination (MPDD). SADV operates in two modes: "In Road Mode" and "Intersection Mode". SNAR make use of optimal paths, which are determined on the basis of graph abstracted from road map. LDU maintains the delay matrix dynamically by measuring the delay of message delivery between static nodes. MPDD helps in multipath routing.

c) Cluster Based Routing

Cluster based routing is preferred in clusters. A group of nodes identifies themselves to be a part of cluster and a n ode is designated as cluster head will broadcast the packet to cluster. Good scalability can be provided for large networks but network delays and overhead are incurred when forming clusters in highly mobile VANET. I n cluster based routing virtual network infrastructure must be created through the clustering of nodes in order to provide scalability. T he various Clusters based routing protocols are COIN and LORA_CBF

d) Broadcast Routing

Broadcast routing is frequently used in VANET for sharing, traffic, weather and emergency, road conditions among vehicles and delivering advertisements and announcements. The various Broadcast routing protocols are BROADCOMM, UMB, V-TRADE, and DV-CAST.

e) Geo Cast Routing

Geo cast routing is basically a location based multicast routing. Its objective is to deliver the packet from source node to all other nodes within a specified geographical region (Zone of Relevance ZOR). In Geo cast routing vehicles outside the ZOR are not alerted to avoid unnecessary hasty reaction. Geo cast is considered as a m ulticast service within a specific geographic region. It normally defines a forwarding zone where it directs the flooding of packets in order to reduce message overhead and network congestion caused by simply flooding packets everywhere. In the destination zone, unicast routing can be used to forward the packet. One pitfall of Geo cast is network partitioning and also unfavorable neighbors, which may hinder the proper forwarding of messages. The various Geo cast routing protocols are IVG, DG-CASTOR and DRG

6. Conclusion

In this section we have reviewed existing routing protocols. Table 2 gives a Comparison of these protocols. Prior forwarding method describes the first routing decision of the protocol when there are packets to be forwarded. In case of Delay Bounded protocols the prior forwarding method is used, whereas in all other routing protocols wireless multi hop method of forwarding is used. Digital map provides street level map and traffic statistics such as traffic density and vehicle speed on road at different times. Digital map is mandatory in case of Some of Cluster Based Routing Protocols. Virtual Infrastructure is created through clustering of nodes in order to provide scalability. Each cluster can have a cluster head, which is responsible for secure communication between inter-cluster and intra cluster coordination in the network. Recovery strategy is used to recover from unfavorable situations. Recovery strategy is the criteria, which is used to judge the performance of protocol.

**Table 2: Comparison of Various Protocols**

| Protocols | Proactive Protocols | Reactive Protocols | Position based Greedy Protocols | Delay Bounded Protocols | Cluster Based Protocols | Broadcast Protocols | Geo cast Protocols |
|---|---|---|---|---|---|---|---|
| **Prior Forwarding Method** | Wire less multi hop Forwarding | Wire less multi hop Forwarding | Heuristic method | Carry & Forward | Wireless Multi hop Forwarding | Wire less multi hop Forwarding | Wire less multi hop Forwarding |
| **Digital Map Requirement** | No | No | No | No | Yes | No | No |
| **Virtual Infrastructure Requirement** | No | No | No | No | Yes | No | No |
| **Realistic Traffic Flow** | Yes | Yes | Yes | No | No | Yes | Yes |
| **Recovery Strategy** | Multi Hop Forwarding | Carry & Forward | Carry & Forward | Multi hop Forwarding | Carry & Forward | Carry & Forward | Flooding |
| **Scenario** | Urban | Urban | Urban | Sparse | Urban | Highway | Highway |